\documentclass[fleqn]{2023SCGE}
\usepackage{natbib}
\usepackage{color}
\usepackage{soul}

\begin{document}

\ensubject{subject}

\ArticleType{Article}
\SpecialTopic{SPECIAL TOPIC: }
\Year{2023}
\Month{January}
\Vol{66}
\No{1}
\DOI{??}
\ArtNo{000000}
\ReceiveDate{xxx, 2024}
\AcceptDate{xxx, 2024}

\title{Quasi-2D Weak Lensing Cosmological Constraints Using the PDF-SYM method}


\author[1]{Zhenjie Liu}{}
\author[1,2]{Jun Zhang}{{betajzhang@sjtu.edu.cn}}
\author[1,3]{Hekun Li}{hekun_lee@sjtu.edu.cn}
\author[1]{Zhi Shen}{}
\author[1]{Cong Liu}{}

\AuthorMark{Liu Z. J.}

\AuthorCitation{Liu, Zhang, et al.}

\address[1]{Department of Astronomy, Shanghai Jiao Tong University, Shanghai 200240, China}
\address[2]{Shanghai Key Laboratory for Particle Physics and Cosmology, Shanghai 200240, China}
\address[3]{Shanghai Astronomical Observatory, Chinese Academy of Sciences, Shanghai 200030, China}


\abstract{
Cosmic shear statistics, such as the two-point correlation function (2PCF), can be evaluated with the PDF-SYM method instead of the traditional weighted-sum approach. It makes use of the full PDF information of the shear estimators, and does not require weightings on the shear estimators, which can in principle introduce additional systematic biases. This work presents our constraints on $S_8$ and $\Omega_m$ from the shear-shear correlations using the PDF-SYM method. The data we use is from the z-band images of the Dark Energy Camera Legacy Survey (DECaLS), which covers about 10000 deg$^2$ with more than 100 million galaxies. The shear catalog is produced by the Fourier\_Quad method, and well tested on the real data itself with the field-distortion effect. Our main approach is called quasi-2D as we do use the photo-$z$ information of each individual galaxy, but without dividing the galaxies into redshift bins. We mainly use galaxy pairs within the redshift interval between 0.2 and 1.3, and the angular range from $4.7$ to $180$ arcmin. Our analysis yields $S_8=0.762 \pm 0.026$ and $\Omega_{\rm m}=0.234 \pm 0.075$, with the baryon effects and the intrinsic alignments included. The results are robust against redshift uncertainties. We check the consistency of our results by deriving the cosmological constaints from auto-correlations of $\gamma_1$ and $\gamma_2$ separately, and find that they are consistent with each other, but the constraints from the $\gamma_1$ component is much weaker than that from $\gamma_2$. It implies a much worse data quality of $\gamma_1$, which is likely due to additional shear uncertainties caused by CCD electronics (according to the survey strategy of DECaLS). We also perform a pure 2D analysis, which gives $S_8=0.81^{+0.03}_{-0.04}$ and $\Omega_{\rm m}=0.25^{+0.06}_{-0.05}$. Our findings demonstrate the potential of the PDF-SYM method for precision cosmology. }



\maketitle


\begin{multicols}{2}
\section{Introduction} \label{sec1}

Weak gravitational lensing is a powerful tool for probing the structure and energy contents of our universe~\citep{BARTELMANN2001,Hoekstra08,Mandelbaum18}. It arises from the gravitational bending of lights by the density fluctuations along line of sights, leading to coherent distortions in galaxy images known as cosmic shear. In 2000, cosmic shear was observed for the first time by four groups independently~\citep{van2000,bacon2000, kaiser2000,wittman2000}, which verified the feasibility and accuracy of the shear measurement method. More recently, there are galaxy surveys with much larger sky coverage to significantly enhance the ability of cosmological constraints, such as the Canada-France-Hawaii Telescope Lensing Survey\footnote{www.cfhtlens.org}, the Dark Energy Survey\footnote{www.darkenergysurvey.org}, the Hyper Suprime-Cam Subaru Strategic Program\footnote{hsc.mtk.nao.ac.jp/ssp/}, and the Kilo-Degree Survey\footnote{kids.strw.leidenuniv.nl}. These surveys have successively published their data and the constraints on the cosmological parameters using cosmic shear~\citep{2013MNRAS.432.2433H,2013MNRAS.430.2200K,2017MNRAS.465.1454H,2017MNRAS.471.4412K,2018PhRvD..98d3528T,2019PASJ...71...43H,2022PhRvD.105b3515S,2022PhRvD.105b3514A}. 
In the near future, a number of large-scale observations will be carried out, such as the Nancy Grace Roman Space Telescope\footnote{www.stsci.edu/roman}, Euclid\footnote{www.euclid-ec.org}, the Large Synoptic Survey Telescope\footnote{www.lsst.org}, and the Chinese Space Station Telescope~\citep{2019ApJ...883..203G}. These projects are expected to bring us unprecedented precision in weak lensing measurements, and therefore deeper insights into cosmology.

Weak lensing provides information about the density fluctuation ($\sigma_8$) and the mean matter density ($\Omega_{\rm m}$) of the universe. Larger values of either of these parameters lead to enhanced lensing effects, exhibiting a strong banana-shaped degeneracy in the $\sigma_8\times\Omega_{\rm m}$ plane. To describe the lensing strength, the combined parameter $S_8=\sigma_8\sqrt{\Omega_{\rm m}/0.3}$ is often used to reduce the degeneracy. 
Currently, the weak lensing data does support the $\Lambda$CDM model. 
However, recent works have found that the weak lensing data prefers a slightly lower $S_8$ compared with the results of CMB data~\citep{2017MNRAS.465.1454H, 2018MNRAS.476.4662V, 2019PASJ...71...43H, 2021A&A...645A.104A, 2022PhRvD.105b3515S, 2022PhRvD.105b3514A}, and possible reasons are studied in a number of literatures~\citep{2017MNRAS.465.2033J, 2017MNRAS.467.3024L, 2022arXiv220207440A}. If such a discrepancy is confirmed, we may need new physics beyond the $\Lambda$CDM model to understand the structure formation of the early and late universe. 

On the other hand, the discrepancy on $S_8$ may be caused by the systematic errors in the weak lensing measurements and analysis. Measuring cosmic shear accurately poses significant challenges given that the shear signals are at one percent level. Technical obstacles include, but not limited to, correcting for the PSF effect~\citep{jarvis2008}, handling CCD effects like Charge-Transfer-Inefficiency~\citep{Massey09} and the brighter-fatter effect~\citep{Antilogus2014}, and addressing photo-z uncertainties~\citep{Pengjie2010,LiuDZ2023}. Astrophysically, uncertainties exist in theories related to intrinsic alignment of galaxies~\citep{2001MNRAS.320L...7C,2018ApJ...853...25W, 2020ApJ...904..135Y,Samuroff2023,Zhang2023,Zhou2023}, baryonic effects~\citep{Schneider2022,Lee2022,Sunseri2023,Chen2023}, and massive neutrinos~\citep{Bayer2022,Lin2022,2022MNRAS.512.3319Z,Abbott2023}. These challenges highlight the complexity involved in lensing measurements.

 In the current situation, we believe it is beneficial to study the consistency of the weak lensing results with more different shear catalogs, or different statistical methods. Recently, we have reported a series of advances of the Fourier\_Quad (called FQ hereafter) shear measurement method~\citep{2015JCAP...01..024Z,2019ApJ...875...48Z}. The shear estimators of FQ are formed by the multipole moments of the galaxy power spectrum. Another recent progress is in the development of a new statistical method called PDF-SYM~\citep{2017ApJ...834....8Z}. It recovers the weak lensing statistics (shear stacking, n-point correlations) by symmetrizing the PDF of the shear estimators with a pseudo signal or the joint PDF of two shear estimators with an assumed correlation strength. The new method does not require weights for the shear estimators, and guarantees that the statistical uncertainties of the final results approach the theoretical lower bound. In this work, based on PDF-SYM, we introduce a new way of studying the shear-shear correlation: quasi-2D analysis. Applying this analysis to the shear catalogue of the FQ method, we obtain substantial constraints on the cosmological parameters.

In \S\ref{sec:pdf}, we introduce the concept of "quasi-2D" shear-shear correlation, and show how to achieve cosmological constraints with the PDF-SYM method. Our main results are given in \S\ref{sec4}, which includes important astrophysical effects such as the intrinsic alignment and the baryonic feedback, as well as the impact of photometric redshift errors. As a consistency check, in \S\ref{sec5}, we show a pure 2D analysis of the shear-shear correlation (i.e., without using individual galaxy's photo-z information) using PDF-SYM. We summarize our findings and provide some discussions in \S\ref{sec6}. 

\section{Quasi-2D shear-shear correlation based on the PDF-SYM Method} 
\label{sec:pdf}
 
PDF-SYM is an unconventional way of evaluating all sorts of weak lensing statistics. It makes use of the full information contained in the PDF of the shear estimators in the case of either shear stacking or shear-shear correlations, as shown in \cite{2017ApJ...834....8Z}. The corresponding shear statistics is evaluated by finding its best value that can bring the PDF back to a symmetric state. It has been demonstrated with the FQ shear estimators that the PDF-SYM method allows the statistical error of the measurement to approach the theoretical limit. The idea has been tested in both simulations and real-data processing with satisfying results. Previous works mainly focus on the shear stacking though, and we in this paper show the performance of PDF-SYM in measuring the shear-shear correlations. 

\subsection{Fourier\_Quad shear estimators}

In the FQ method, there are five components from each galaxy that are relevant to shear measurement: $G_1$, $G_2$, $N$, $U$, and $V$, where $G_i$ is similar to the ellipticity components $e_i$, and $N$ is a normalization factor. $U$ and $V$ are the additional correction terms. They are defined as:
\begin{eqnarray}
&&G_1=-\frac{1}{2} \int d^2 \vec{k}\left(k_x^2-k_y^2\right) T(\vec{k}) M(\vec{k})\\ \nonumber
&&G_2=-\int d^2 \vec{k} k_x k_y T(\vec{k}) M(\vec{k})\\ \nonumber
&&N=\int d^2 \vec{k}\left(k^2-\frac{\beta^2}{2} k^4\right) T(\vec{k}) M(\vec{k})\\ \nonumber
&&U=-\frac{\beta^2}{2} \int d^2 \vec{k}\left(k_x^4-6 k_x^2 k_y^2+k_y^4\right) T(\vec{k}) M(\vec{k})\\ \nonumber
&&V=-2 \beta^2 \int d^2 \vec{k}\left(k_x^3 k_y-k_x k_y^3\right) T(\vec{k}) M(\vec{k})
\end{eqnarray}
in which $M(\vec{k})$ represents the power spectrum of galaxy images after removing background and Poisson noise, and the factor $T(\vec{k})$ is used to transform the PSF to the desired isotropic Gaussian form. $\beta$ is the scale radius of the Gaussian function (see \cite{2015JCAP...01..024Z} for more details).
The shear values can be simply computed by averaging the first three shear estimators,
\begin{equation}
\frac{\left\langle G_i\right\rangle}{\langle N\rangle}=g_i+O\left\langle g^3\right\rangle
\label{ave}
\end{equation}
in which $g_i$ is the corresponding component of the reduced shear. It is noted that this averaging method is significantly influenced by very bright galaxies, even though the majority of galaxies are faint. Zhang et al. \cite{2017ApJ...834....8Z} proposed a novel statistical approach, the PDF-SYM method, as an alternative to the averaging method to mitigate this issue. 

\subsection{PDF-SYM for shear stacking}
The proposal of PDF-SYM is about observing the symmetry properties of the distributions of $\hat{G}_1(=G_1-\hat{g}_1(N+U))$ and $\hat{G}_2(=G_2-\hat{g}_2(N-U))$, where $\hat{g}_1$ and $\hat{g}_2$ are shear value we guess in the symmetrization. As demonstrated in \cite{2017ApJ...834....8Z}, if $\hat{g}_1=g_1$ and $\hat{g}_2=g_2$, the PDFs of $\hat{G}_1$ and $\hat{G}_2$ are symmetric. We can therefore recover the shear values by finding $\hat{g}_1$ and $\hat{g}_2$ that can best symmetrize the PDFs of $\hat{G}_i$. The PDF-SYM method allows the statistical uncertainty of the result to approach the theoretical lower limit (the Cramer-Rao bound). The quantity $V$ is reserved for calculating the value of $U$ under coordinate rotation, as $(U,V)$ forms a pair of spin-4 quantity.

For convenience, let us firstly define the completely symmetrized version of the shear estimators as:
\begin{equation}
G^{{\rm S}}_i=G_i-g_i B_i
\end{equation}
where $g_i$ is the true shear signal, and $B_i$ is defined as:
\begin{equation}
  B_i= \left \{
  \begin{array}{rcl}
    N+U & & (i=1)\\
    N-U & & (i=2)\\
  \end{array} \right .
\end{equation}
Note that $G^{{\rm S}}_i$ is not known, but it is defined mainly for the fact that its PDF is symmetric with respect to zero. We guess a pseudo-shear $\hat{g}_i$ to modify $G_i$, and the modified shear estimator is denoted as $\hat{G}_i$, i.e., 
\begin{equation}
\label{Ghat}
\hat{G}_i=G_i-\hat{g}_iB_i=G^{\rm S}_i+(g_i-\hat{g}_i)B_i.
\end{equation}
Eq.(\ref{Ghat}) clearly shows that the PDF of $\hat{G}_i$ is only symmetric when $\hat{g}_i=g_i$, because the PDF of $B_i$ is generally not symmetric at all. Therefore, shear value can be estimated by finding the value of $\hat{g}_i$ so that the PDF of $\hat{G}_i$ can be best symmetrized with respect to zero. Technically, this is done by setting up bins on the two sides of zero in symmetric positions. The galaxy number in each bin is counted as one varies the value of $\hat{g}_i$, and a $\chi^2$ as a function of $\hat{g}_i$ is formed to characterize and quantify the level that the PDF of $\hat{G}_i$ deviates from the symmetric state. More details can be found out in section 3.2 of \cite{2017ApJ...834....8Z}. The performance of PDF-SYM method on shear stacking has been demonstrated in the studies of galaxy-galaxy lensing \citep{Fong2022,Wang2022,Xu2023} and shear map reconstruction \citep{Wang2023}.

\subsection{PDF-SYM for shear-shear correlation}
To measure the shear-shear correlation, one can form the joint PDF of the modified shear estimators $P(\hat{G} _i,\hat{G}_i^\prime)$ from galaxy pairs that share the same correlation strength. Based on the derivation in eqs.(38-42) of Ref.\cite{2017ApJ...834....8Z}, the joint distribution $P(\hat{G} _i,\hat{G}_i^\prime)$ can be expressed as:
\begin{eqnarray}
\label{Pgg}
P(\hat{G} _i,\hat{G}_i^\prime) &= &\int dB \int dB^\prime {\textbf [}P_{\rm S}(\hat{G}_i,B,\hat{G}_i^{\prime},B^\prime) \nonumber\\
&+& \frac{1}{2}(\langle g_i^2 \rangle + \langle \hat{g}_i^2 \rangle)(B^2 \partial_{\hat{G}_i}^2 P_{\rm S} + B^{\prime 2} \partial_{\hat{G}_i^\prime}^2 P_{\rm S}) \nonumber\\
&+&(\langle g_i g_i^\prime \rangle + \langle \hat{g}_i \hat{g}_i^\prime \rangle) BB^\prime \partial_{\hat{G}_i} \partial_{\hat{G}_i^\prime} P_{\rm S}{\textbf ]}.
\end{eqnarray} 
in which $P_{\rm S}$ is an even function with respect to its arguments $\hat{G}_i$ and $\hat{G}_i^{\prime}$, referring to the PDF of the shear estimators with the lensing effect being properly corrected. The calculation there assumes that the shear signals are small, and therefore only leading terms are included. The symmetry of the joint PDF can be quantified by forming the following quantity from the four quadrants given the properties of $P_{\rm S}$:
\begin{eqnarray}
\label{sym1}
&&P(\hat{G} _i,  \hat{G}_i')+P(- \hat{G} _i,- \hat{G}_i') -P(- \hat{G} _i,\hat{G}_i')-P(\hat{G} _i,-\hat{G}_i') \nonumber \\
&&= \int dB \int dB' (\langle g_ig_i'\rangle+\langle \hat{g}_i\hat{g}_i'\rangle) BB' \partial_{\hat{G}_i} \partial_{\hat{G}_i'} P_{\rm S}.
\end{eqnarray}
From this formula, it is clear that when $\langle \hat{g}_i\hat{g}_i'\rangle=-\langle g_ig_i'\rangle$, the resulting joint PDF would become symmetric in the four quadrants. In practice, the terms on the left of eq.(\ref{sym1}) would be grouped into a few bins that are symmetrically distributed in the four quadrants. The number of galaxy pairs in each bin is labelled as $n_{ij}$, in which $i$ and $j$ are the bin indices, indicating the ranges of $\hat{G}$ and $\hat{G}'$. One can then constrain the value of $\hat{\xi}$ by minimizing the $\chi^2$ defined as:
\begin{equation}
\label{chi21}
\chi^2 = \frac{1}{2} \sum_{i,j > 0}\frac{(n_{i,j}+n_{-i,-j}-n_{-i,j}-n_{i,-j})^2}{n_{i,j}+n_{-i,-j}+n_{-i,j}+n_{i,-j}}
\end{equation} 
Note that negative value of the bin index (-i or -j) refers to the case that the bin is at the negative side of zero.

To symmetrize the PDF, Zhang et al.~\citep{2017ApJ...834....8Z} generate the pseudo-shear pair $\hat{g}_i$ and $\hat{g}_i'$ for each galaxy pair using a joint Gaussian distribution with predetermined correlation $\langle \hat{g}_i\hat{g}_i'\rangle$. This procedure can however be simplified: for an assumed correlation $\hat{\xi}(=\langle \hat{g}_i\hat{g}_i'\rangle)$, if $\hat{\xi} \ge 0$, we can simply set $\hat{g} = \hat{g}'={\hat{\xi}}^{1/2}$, and otherwise ($\hat{\xi} < 0$), we choose $\hat{g} = -\hat{g}' =(-\hat{\xi})^{1/2}$.

\subsection{Quasi-2D analysis} 
\begin{figure*}[ht!]
\centering
\includegraphics[width=0.7\textwidth]{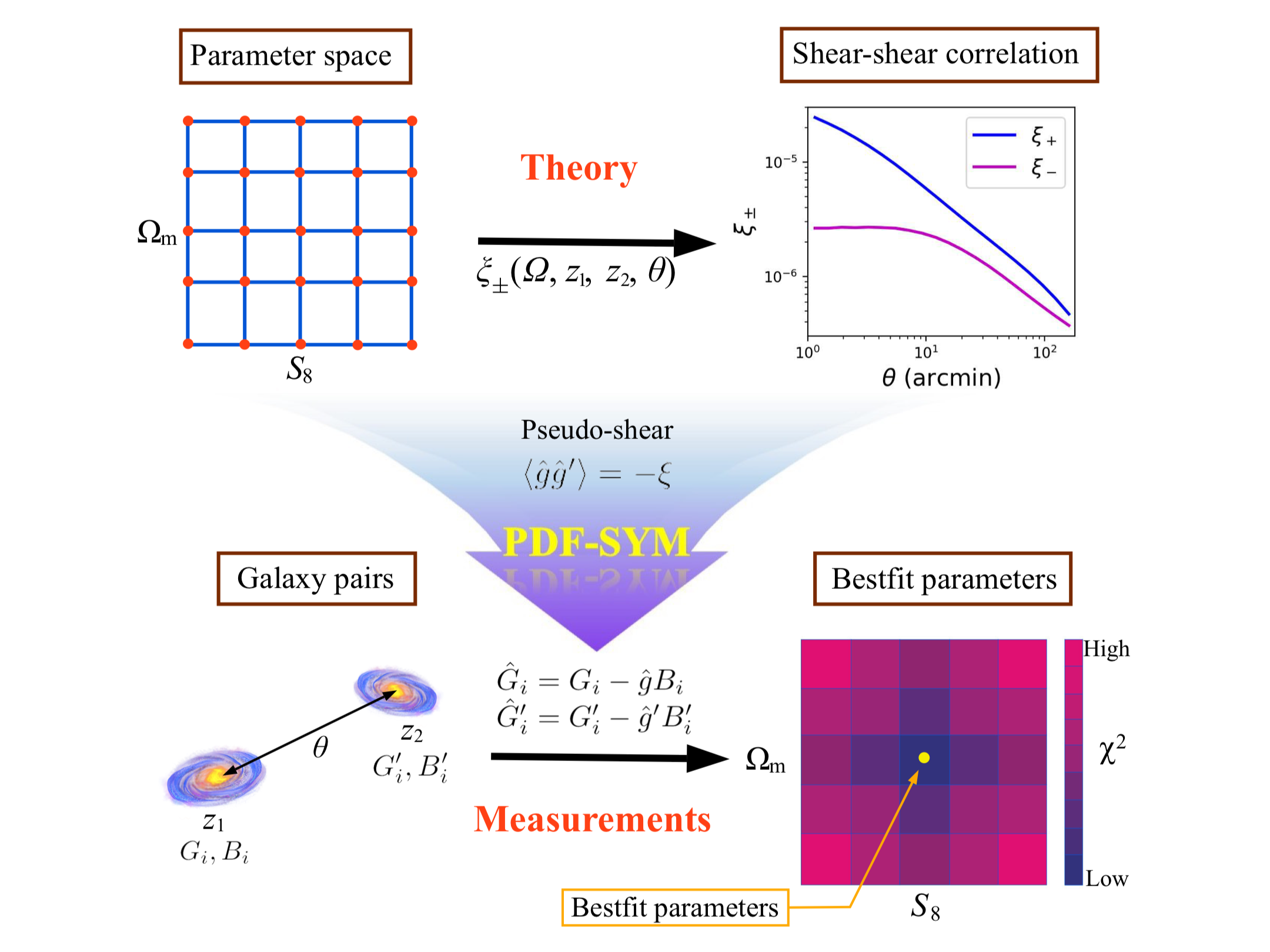}
\caption{The flowchart of the PDF-SYM method for finding the best-fit parameters. Here $S_8$ and $\Omega_{\rm m}$ are  two example parameters. The upper panel shows the correlation function predictions in different cosmological model. We then use the PDF-SYM method to adjust the shear estimates ($G_i$) for each galaxy pair in our measurements, ultimately resulting in a $\chi^2$ map in the parameter space. Finally, we can identify the best-fit parameters corresponding to the minimum $\chi^2$.} 
\label{bestfit}
\end{figure*}

When the galaxies span a significant range in redshift, the shear correlation $\xi$ between different pairs of galaxies is no longer the same value. Stacking all galaxy pairs across different redshifts and using a single $\hat{\xi}$ to symmetrize the overall PDF is a crude approximation (actually it is the 2D analysis, which we discuss in Sec.\ref{sec5} for a comparison). In quasi-2D, we assign a value of $\hat{\xi}$ to each pair of galaxies according to their redshifts (and also their angular separation). The joint PDF should now be integrated over the redshift distribution as:
\begin{eqnarray}\label{sym2}
&&P(\hat{G} _i,  \hat{G}_i^\prime)+P(- \hat{G} _i,- \hat{G}_i^\prime) -P(- \hat{G} _i,\hat{G}_i^\prime)-P(\hat{G} _i,-\hat{G}_i^\prime)\nonumber\\
&&\approx \int dz_1 \int dz_2 [\langle g_i(z_1) g_i^\prime(z_2) \rangle 
 + \hat{\xi} (z_1,z_2,\hat{\Omega}_1,\hat{\Omega}_2,...)] \nonumber\\
&&\times\int dB \int dB'\cdot BB' \partial_{\hat{G}_i} \partial_{\hat{G}_i'} P_{\rm S}(\hat{G}_i,B,\hat{G}_i',B',z,z')
\end{eqnarray}
where $\Omega_1,\Omega_2,...$ refer to the set of cosmological parameters of our interest, and $\hat{\Omega}_1,\hat{\Omega}_2,...$ are their assumed values respectively. Note that we have assumed that the galaxy pairs are all within a certain angular range, and therefore the angular dependence is omitted in eq.(\ref{sym2}). When the assumed parameters approach their real values, the $\chi^2$ in eq.(\ref{chi21}) should reach its minimum value (assuming the cosmological model is correct). This fact allows us to constrain the cosmological parameters, which is the main proposal of this paper. Note that in doing so, we still need to use the galaxy redshift information. This is why the method is called quasi-2D. As discussed later in the paper, the quasi-2D method is more immune to the photo-z errors than regular tomography. A schematic view of this new approach is shown in Figure \ref{bestfit}.

To quantify the symmetry of the PDF, we adopt the definition of $\chi^2$ in eq.(\ref{chi21}), and sum up the $\chi^2$ from different angular bins to form the final one, as we expect that the correct cosmological parameters should symmetrize the PDFs of all the angular bins simultaneously. 
To determine the best-fit parameters, we utilized univariate polynomial fitting to find minimum $\chi^2$ within the parameter space. Taking the example of two free parameters, $S_8$ and $\Omega_{\rm m}$,  we first minimize $\chi^2(S_8,\Omega_{\rm m})$ by varying $S_8$ while fixing $\Omega_{\rm m}$ at different values, yielding best-fit $S_8^{\rm f1}(\Omega_{\rm m})$ and its  corresponding $\chi^2_{\rm min}(S_8^{\rm f1}(\Omega_{\rm m}), \Omega_{\rm m})$. Subsequently, we vary $\Omega_{\rm m}$ to fit $\chi^2_{\rm min}(S_8^{\rm f1}(\Omega_{\rm m}), \Omega_{\rm m})$ and determined the best-fit $\Omega_{\rm m}^{\rm bf}$. Finally, we interpolated this result into the function $S_8^{\rm f1}(\Omega_{\rm m}^{\rm bf})$ to derive $S_8^{\rm bf}$, so that we have best-fit parameters $S_8^{\rm bf}$ and $\Omega_{\rm m}^{\rm bf}$. For more free parameters, we repeat the above steps for each parameter to obtain their best-fit values.

We use the Jackknife approach to get the final constraints of parameters. The observed area is divided into $N_{\rm J}=200$ subregions using the K-means algorithm in Scikit-learn \citep{scikit-learn}. Then, we calculate the best-fit parameters using the PDF-SYM method for each Jackknife sample, denoted as $\Omega_{\rm J}$, and we can compute their mean values and covariance matrix by
\begin{equation}\label{mean}
 \overline{\Omega}_j  = \frac{1}{N_{\rm J}}\sum_{i=1}^{N_{\rm J}} \Omega_{{\rm J},j}^i
\end{equation}
\begin{equation}\label{cov}
{\rm Cov}(\Omega_j, \Omega_k) = \frac{N_{\rm J}-1}{N_{\rm J}}\sum_{i=1}^{N_{\rm J}} (\Omega_{{\rm J},j}^i - \overline{\Omega}_j )(\Omega_{{\rm J},k}^i -\overline{\Omega}_k)
\end{equation}
where $j$ and $k$ are the indexes for the parameters and $i$ is the index for the Jackknife samples. Hence $\overline{\Omega}_j $ is the final value of a parameter and its $1\sigma$ error is calculated by $\sqrt{{\rm Cov}(\Omega_j, \Omega_j)}$.

\section{data and results}\label{sec4}

In this section, we give a brief introduction of the shear catalog used in our measurements, the cosmological model and shear-shear correlation calculation for the cosmological constraints. We present our main cosmological results, and assess their robustness by examining the impact of various shear components of shear-shear correlation, followed by an examination of the effects of photo-z errors. We consider the contributions from the intrinsic alignment and the corrections to the density power spectrum from the baryonic effect in all of our tests.

\subsection{Data}\label{sec:data}
\begin{figure}[H]
\centering
\includegraphics[width=0.43\textwidth]{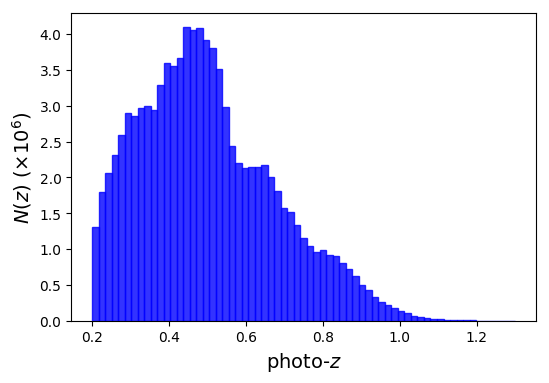}
\caption{The photo-$z$ distribution of galaxies used in our analysis.}  \label{fig:zpdf}
\end{figure}

Our shear catalog is obtained from the imaging data of the Dark Energy Camera Legacy Survey (DECaLS) processed by the FQ shear measurement pipeline \citep{2022arXiv220602434Z}. DECaLS, one of the three public projects in Dark Energy Spectroscopic Instrument (DESI) Legacy Imaging Surveys \citep{2019AJ....157..168D}, aims to select target for DESI under poor seeing conditions ($\sim$ 1.5"). It covers about 10000 deg$^2$ sky in $g$, $r$ and $z$ bands.  
Our shear catalog contains the photo-$z$ measured by Zhou
et al. \cite{2021MNRAS.501.3309Z}. According to Zhang et al. \cite{2022arXiv220602434Z}, the z-band shear catalog has the best quality, which has the lowest multiplicative and additive biases. We therefore only use the z-band shear catalog in our analysis, with typical galaxy number density about 3 - 5 per square arcmin. We select galaxies from the z-band with photo-$z$ ranging from 0.2 to 1.3, and $\nu_F$ (a definition of the signal-to-noise-ratio in Fourier space \cite{Li_2021}) larger than 4. We also remove the shear measurements from the edges of the exposures with $\vert g_{f1,f2}\vert >0.0015$ (where $g_{f1}$ and $g_{f2}$ refer to the two ellipticity components of the field distortion), and those from problematic chips. Besides, we also remove galaxies with magnitudes (mag) larger than 21, whose photo-$z$ are more difficult to measure accurately due to their low brightness (Li et al. in prep.). Finally, we get about $9.7\times 10^7$ galaxy images. Figure \ref{fig:zpdf} shows the photo-$z$ distribution of the selected galaxies. Our 2PCFs measurement covers the angular distance from $\theta_{\rm min} = 1$ arcmin to $\theta_{\rm max} = 180$ arcmin, evenly divided into 20 bins in log-space. It is worth noting that we only use galaxy pairs from different exposures, thus avoiding additional shear correlations caused by systematics within the same exposure, such as PSF leakage \citep{Jarvis_2016,Lu_2017}.

\subsection{Theory and Model}\label{sec3}

Shear-shear correlation is usually measured with the tangential and cross shear components ($\gamma_t$ and $\gamma_\times$) defined in the coordinate connecting the two galaxies. The shear 2PCF $\xi_\pm$ is defined as
\begin{eqnarray}
\label{eq:xitt}
\xi_\pm(z_1,z_2,\Delta\vec{\theta}) 
&=& \langle\gamma_t(z_1,\vec{\theta}) \gamma_t(z_2,\vec{\theta}+\Delta\vec{\theta})\rangle \\ \nonumber
&\pm& \langle\gamma_{\times}(z_1,\vec{\theta}) \gamma_{\times}(z_2,\vec{\theta}+\Delta\vec{\theta})\rangle.
\end{eqnarray}
In addition to this traditional choice, we consider another two correlation functions defined with only $\gamma_1$ or $\gamma_2$ respectively, i.e., 
\begin{equation}
\label{eq11}
\xi_{ii}(z_1,z_2,\Delta\vec{\theta}) 
=\langle\gamma_i(z_1,\vec{\theta}) \gamma_i(z_2,\vec{\theta}+\Delta\vec{\theta})\rangle,
\end{equation}
with $i=1,2$. Note that the CCD images of the DECaLS exposures all line up with the equatorial direction, which is also the coordinate direction for the definition of the shear components in our DECaLS shear catalog. Separate studies of the correlation functions of $\gamma_1$ and $\gamma_2$ therefore allow us to find out if there are residual shear measurement errors, particularly those related to the charge transfer along the pixel readout direction, which can in principle cause significant systematic effects in the correlation function measurement. As shown later in the paper, the cosmological constraints from $\xi_{11}$ is indeed much worse than those from $\xi_{22}$. Theoretically, we expect $\xi_{11} = \xi_{22}=\xi_+/2$.

Including the intrinsic alignment, the shear-shear correlation can be written as
\begin{equation}
\label{shear_shear}
 \xi_{\pm}= \xi_{\rm GG \pm}+\xi_{\mathrm{II} \pm}+\xi_{ \mathrm{GI} \pm},
\end{equation}
in which the subindex "I" stands for the intrinsic galaxy shape, and "G" represents the cosmic shear. The first term is the shear-shear correlation. The second term is the auto-correlation of the intrinsic alignment, which mainly comes from the interaction of nearby galaxies. The last term $\xi_{ \mathrm{GI}}$ represents the correlation between the intrinsic ellipticity and the background shear signal, which is due to the connection between the galaxy intrinsic shape and the foreground matter field. Following  Heymans et al. \cite{2013MNRAS.432.2433H}, given a group of galaxy pairs with the angular distance $\theta$ and the normalized redshift distributions $n^{i}(z)$ and $n^{j}(z)$, we model the shear correlation function as:

\begin{equation}\label{xipm}
 \xi^{ij}_\pm(\theta) = \int_0^\infty \frac{d\ell}{2\pi} \ \ell \ C^{ij}(\ell) \ J_{\nu} (\ell \theta),
\end{equation}
\begin{equation}\label{cl}
C^{ij}(\ell) = \int_0^\infty dz \  \frac{W^i(z) W^j(z)}{\chi(z)^2} P_\delta \left(\frac{\ell}{\chi(z)}, z\right)
\end{equation}
where $\chi (z)$ represents the comoving radial distance at the redshift $z$. $P_\delta$ represents the nonlinear matter power spectrum, calculated by the Core Cosmology Library \cite{2019ApJS..242....2C}, and the nonlinear evolution is described by the halofit model \citep{10.1046/j.1365-8711.2003.06503.x,2012ApJ...761..152T}. $\ J_{\nu}$ is the Bessel function of the first kind, and $\nu$ is 0 for $\xi_+$ and 4 for $\xi_-$. $W(z)$ is the kernel including the contributions from both lensing and the intrinsic alignment:
\begin{equation}\label{window}
W^{i}(z) = W^{i}_{\rm G}(z) + W^{i}_{\rm I}(z).
\end{equation}
The lensing kernel $W^{i}_{\rm G}(z)$ takes the form of
\begin{equation}\label{windowG}
W^{i}_{\rm G}(z) = \frac{3}{2}\Omega_m \ \frac{H_0^2}{c^2} \frac{\chi(z)}{a(z)} \int_z^\infty dz^{\prime} n^{i}(z^{\prime}) \ \frac{\chi(z^{\prime})-\chi(z)}{\chi(z^{\prime})},
\end{equation}
with $a(z)$ is the scale factor, $H_0$ is Hubble constant, and $c$ is the speed of light. Note that we assume a flat universe in this work for simplicity. For the IA kernel, we adopt the Nonlinear Alignment (NLA) Model \citep{2007NJPh....9..444B}. The NLA model believe that the average intrinsic ellipticity of galaxies is directly proportional to the gravitational potential at the time of galaxy formation, and it incorporates non-linear matter power spectrum. The kernel has the following form:
\begin{equation}\label{windowI}
W^{i}_\mathrm{I}(z) = -\frac{A_{\rm IA} C_1 \rho_c \Omega_\mathrm m}{D(z)}  \ n^{i}(z),
\end{equation}
where $D(z)$ is the normalized growth factor, $\rho_c$ is the critical density and $A_{\rm IA}$ is a free parameter describing the amplitude of IA. $C_1$ is a normalization constant that can be set as $C_1=5 \times 10^{-14} h^{-2} M_\odot^{-1} {\rm Mpc}^3$ to match the observational results in \citep{2002MNRAS.333..501B} so that the fiducial value of $A_{\rm IA}$ is 1. Combining eq.(\ref{xipm}), (\ref{cl}) and (\ref{windowG}), we have the theoretical model of $\xi_\pm$. 

In our case, we assume that the redshifts of each galaxy pair are known, say $z_1$ and $z_2$, we therefore use the Dirac delta function $\delta_D(z_1)$ and $\delta_D(z_2)$ to describe their redshift distributions. This is a fine choice for the lensing kernel defined in eq.(\ref{windowG}), but not so in eq.(\ref{windowI}) as it leads to infinity. This is the result of the limber's approximation one typically adopts in the derivation of the weak lensing formalism, which requires a smooth redshift kernel. For this reason, we set $n^{i}(z)$ as a very narrow Gaussian distribution with a standard deviation of $0.005(1+z)$ in eq.(\ref{windowI}) for intrinsic alignment. One can think of this as if we are using the average IA of an ensemble of neighboring (in redshift) galaxy pairs to represent the IA of a single galaxy pair, which should be an acceptable choice. 

The matter power spectrum is modified by the baryonic effect, which is described by the baryonic correction model (BCM) \cite{2015JCAP...12..049S} in this work. It relates the total matter density field to the gas and stars, and parameterizes the modifications on the power spectrum. The two vital parameters in the BCM model are the mass fraction of ejected gas ($M_c$) and the ejection radius (which depends on the parameter $\eta_b$). For our primary analyses, we adopt the fiducial values from Schneider and Teyssier \cite{2015JCAP...12..049S}: $M_c=1.2 \times 10^{14} M_\odot /h$ and $\eta_b=0.5$, which are matched by simulations and observations in their work. Therefore, in our theoretical predictions, we integrate the BCM and NLA models, and constrain the three free parameters $A_{\rm IA}$, $S_8$, and $\Omega_{\rm m}$. The other cosmological parameters are fixed on the values from Planck~\citep{2020A&A...641A...6P}.

\subsection{The Baseline Constraints}\label{base_res}

\begin{figure}[H]
\centering
\includegraphics[width=0.48\textwidth]{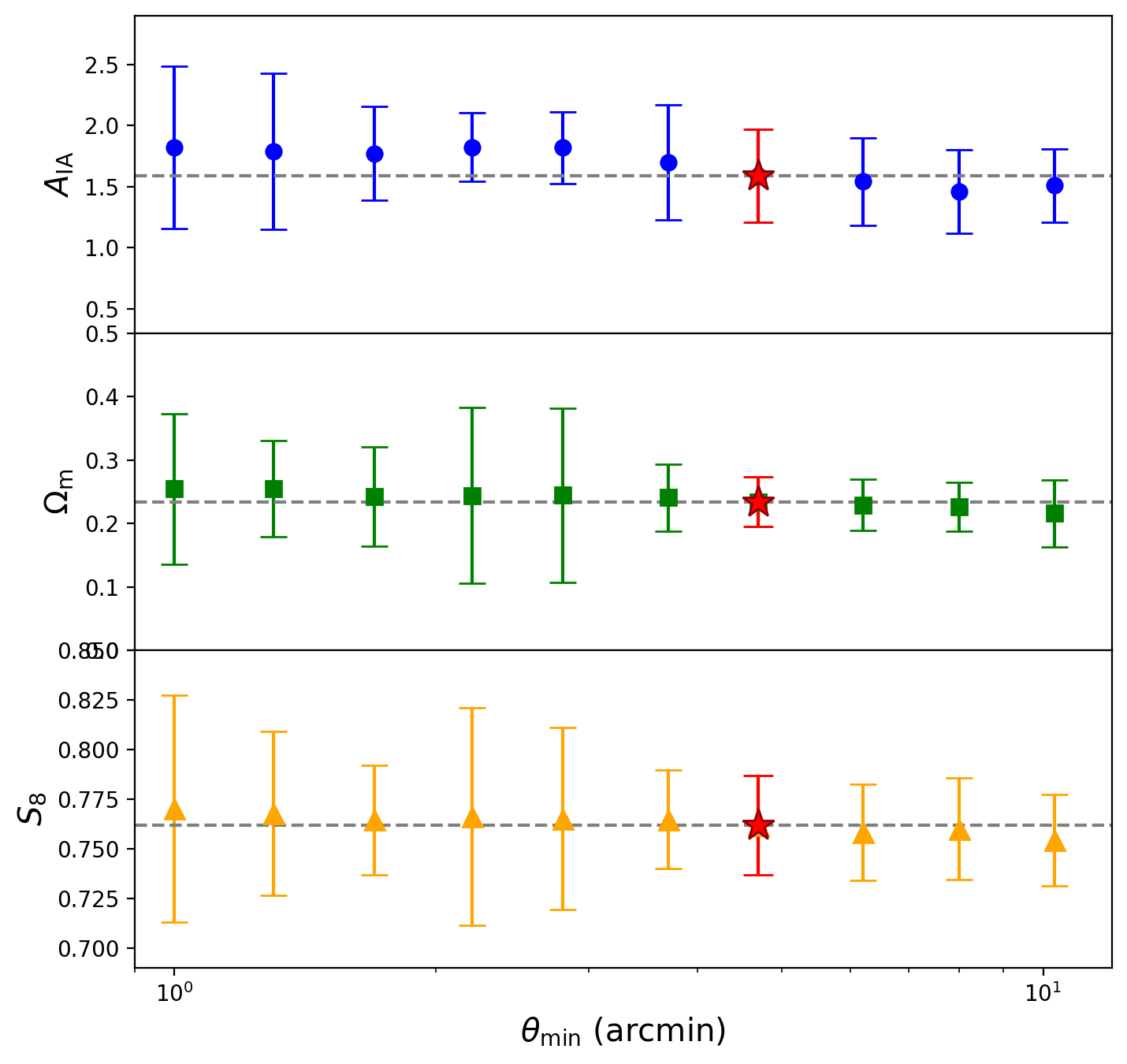}
\caption{The constraint of $A_{\rm IA}$, $\Omega_{\rm m}$ and $S_8$ using quasi-2D analysis with both $\xi_{\gamma_t \gamma_t}$ and $\xi_{\gamma_\times \gamma_\times}$ components. The minimum angle $\theta_{\rm min}$ of galaxy pairs is varied, but the largest scale cut $\theta_{\rm max}$ is fixed at 180 arcmin for each point. The pentagram shows our final choice for the baseline results.  \label{fig:AIAthmin}}
\end{figure}

In this section, we show the parameter constraints obtained by the quasi-2D analysis using both $\xi_{\gamma_t \gamma_t}$ and $\xi_{\gamma_\times \gamma_\times}$. Since the baryonic effect and the intrinsic alignment are more pronounced on small scales, we vary the minimum angle $\theta_{\rm min}$ of galaxy pairs ($\theta_{\rm max}$ remains fixed at 180 arcmin) to find out how the cosmological constraints vary in Figure \ref{fig:AIAthmin}. The results fluctuate slightly on the smaller scales as we gradually remove the bins of small angular separations. It becomes stable when $\theta_{\rm min}\gtrsim 4$ arcmin. Contrary to the usual expectation that a smaller $\theta_{\rm min}$  tightens parameter constraints by including more data, our findings suggest otherwise. We suspect this is due to imperfect modeling of small-scale phenomena, influenced significantly by baryonic effects or intrinsic alignments. Despite these models offer reasonable predictions after redshift integration, inaccuracies in predicting individual galaxy pairs may lead to a more dispersed PDF of shear estimators, consequently resulting in larger error bars.

By setting $\theta_{\rm min}= 4.7$ arcmin, we get our baseline constraints on the parameters with 1 $\sigma$ errors (marked in the Figure \ref{fig:AIAthmin} with a pentagram):
$S_8 = 0.762 \pm 0.026$, $\Omega_{\rm m} = 0.234 \pm 0.075$ and $A_{\rm IA}= 1.59 \pm 0.73$. Despite not employing redshift binning, our optimized methodology and the ample dataset have still yielded substantial constraints. For the intrinsic alignment, since the GI term of IA suppresses the 2PCFs, and the II term elevates them, the effects of GI and II are cancelled to some extent in traditional 2D lensing, and the results are therefore not so sensitive to IA. But in our quasi-2D method, we can constrain $A_{\rm IA}$ well and its result is consistent with the fiducial value, which is also similar to the results from other works~\citep{2017MNRAS.465.1454H,2020PASJ...72...16H,2018MNRAS.476.4662V}.

\subsection{Consistency Test with Different Shear Components}\label{internal}

\begin{figure}[H]
\centering
\includegraphics[width=0.48\textwidth]{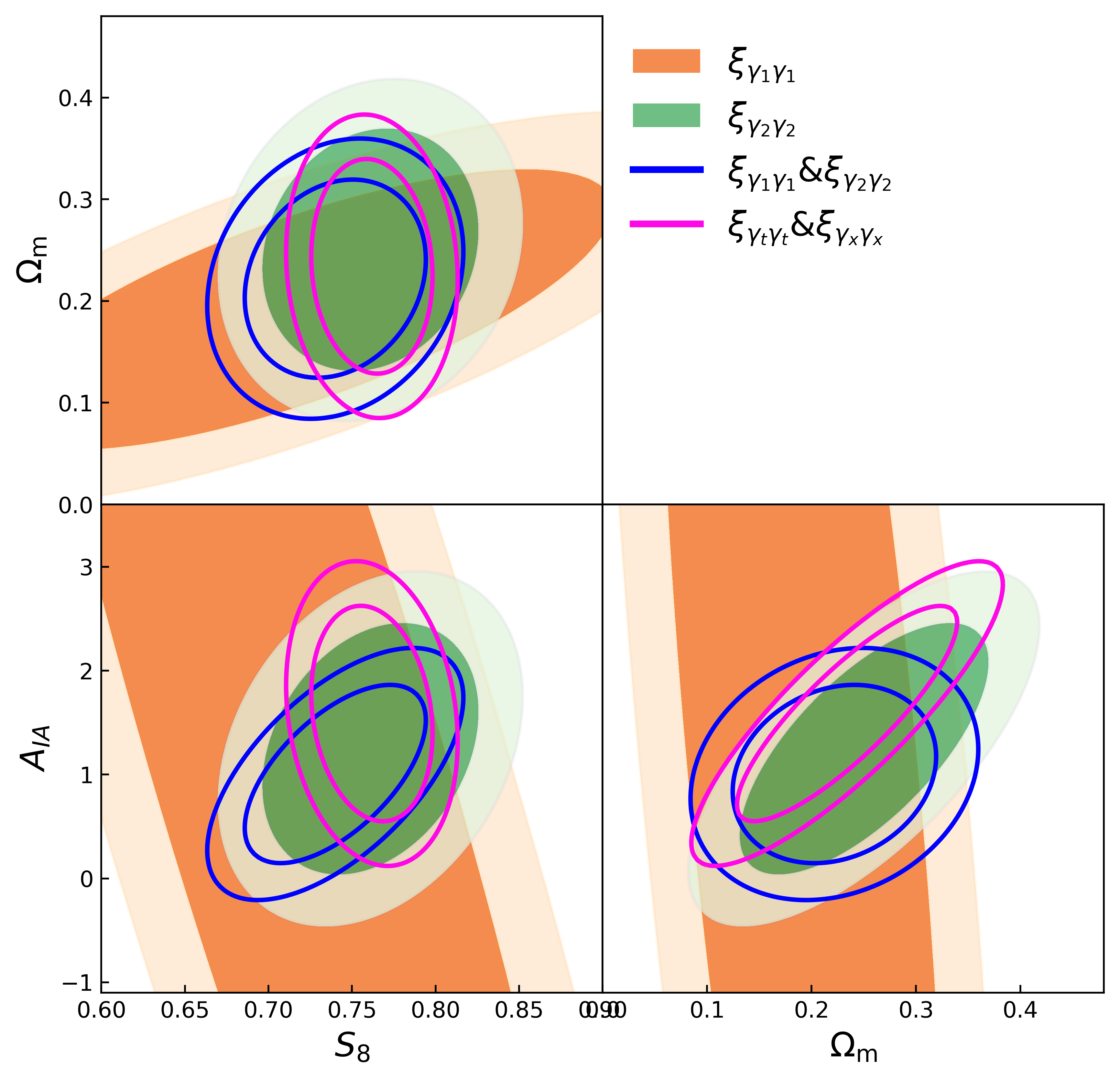}
\caption{The 68\% and 95\% confidence level contour plots of three cosmological parameters: $S_8$, $\Omega_{\rm m}$, and $A_{\rm IA}$, using our quasi-2D analysis. The orange and green contours are calculated using $\xi_{\gamma_1 \gamma_1}$ and $\xi_{\gamma_2 \gamma_2}$ components respectively, while the blue curves represent the constraints obtained by combining both components. The pink curve represents the baseline results obtained by using both $\xi_{\gamma_t \gamma_t}$ and $\xi_{\gamma_\times \gamma_\times}$ components. \label{fig:AIAcontour}}
\end{figure}

In this part, we examine the internal consistency with different shear components. Since the direction of $\gamma_1$ is aligned with the orientation of CCD pixels, it is more likely to be affected by the CCD electronics, e.g., the Charge Transfer Inefficiency~\citep{Massey09}. Hence, it is interesting to examine the consistency of the results using $\gamma_1$ and $\gamma_2$ separately. Figure \ref{fig:AIAcontour} shows the 68\% and 95\% confidence level (CL) contour plots achieved using $\xi_{\gamma_1 \gamma_1}$ and $\xi_{\gamma_2 \gamma_2}$, both jointly and separately. For comparison, we also show the constraints from $\xi_{\gamma_t \gamma_t}$ and $\xi_{\gamma_\times \gamma_\times}$ jointly. The contours in the figure are drawn using the mean values in eq.(\ref{mean}) and covariance matrices in eq.(\ref{cov}) of the parameters, assuming that the results of the parameters conform to an ideal Gaussian distribution, and are therefore ellipse-like. The numerical results are listed on the first four rows of Table \ref{tab:allresults}. It is evident that $\xi_{\gamma_1 \gamma_1}$ itself does not constrain $S_8$ and $A_{\rm IA}$ very well (orange contours), and the value of $\Omega_{\rm m}$ is somewhat lower than those of other cases. In contrast, the parameter constraints from $\xi_{\gamma_2 \gamma_2}$ only (green contours) turn out to be more stringent and reasonable. In principle, $\xi_{\gamma_1 \gamma_1}$ and $\xi_{\gamma_2 \gamma_2}$ should have comparable abilities in constraining the cosmological parameters. However, our results indicate that this is not the case. The quality of the $\gamma_1$ shear catalog seems to be worse than that of $\gamma_2$.

\begin{table*} 
\centering  
\caption{Summary of constraints of three parameters, $S_8$, $\Omega_{\rm m}$ and $A_{\rm IA}$, at 1$\sigma$ errors in different settings. "Quasi-2D" represents the results from the quasi-2D analysis, while "2D" indicates the use of the pure 2D method. 
“Worse photo-$z$” refers to the case with extra photo-z errors added to the galaxies in the shear catalogue. A visual comparison of $S_8$ can be seen in Figure \ref{diff_S8}.} \label{tab:allresults}
\begin{tabular}{lccc}\toprule[0.65pt]
\hline  
Setups & $S_8$ & $\Omega_{\rm m}$& $A_{\rm IA}$\\
[2.4pt]
\hline
Quasi-2D $\xi_{\gamma_t \gamma_t} \& \xi_{\gamma_\times \gamma_\times}$ & $\mathbf{0.762 \pm 0.026}$ &  $\mathbf{0.23 \pm 0.07}$ & $\mathbf{1.59 \pm 0.73}$\\
Quasi-2D $\xi_{\gamma_1 \gamma_1} \& \xi_{\gamma_2 \gamma_2}$ &$0.740 \pm 0.038$ & $0.22 \pm 0.07$ & $1.00 \pm 0.60$ \\ 
Quasi-2D $\xi_{\gamma_1 \gamma_1} $ &$0.719 \pm 0.132$ & $0.19 \pm 0.10$ & $1.05 \pm 6.43$ \\ 
Quasi-2D $\xi_{\gamma_2 \gamma_2}$ &$0.761 \pm 0.046$ & $0.25 \pm 0.08$ & $1.25 \pm 0.85$ \\ 
Worse photo-$z$ &$0.761 \pm 0.031$ & $0.24 \pm 0.06$ & $2.37 \pm 1.04$ \\ 
2D $\xi_{\pm}$ & $0.81^{+0.03}_{-0.04} $ &  $0.25^{+0.06}_{-0.05}$ & $2.47^{+1.35}_{-1.16}$ \\
2D $\xi_{\gamma_1 \gamma_1} \& \xi_{\gamma_2 \gamma_2}$  & $0.74^{+0.05}_{-0.06} $ &  $0.17^{+0.06}_{-0.05}$ & $2.50^{+2.67}_{-2.01}$ \\
2D $\xi_{\gamma_1 \gamma_1}$  & $0.73^{+0.06}_{-0.07} $ &  $0.17^{+0.08}_{-0.05}$ & $2.88^{+2.36}_{-2.16}$ \\
2D $\xi_{\gamma_2 \gamma_2}$  & $0.76^{+0.06}_{-0.07} $ &  $0.22^{+0.10}_{-0.07}$ & $2.65^{+2.52}_{-2.13}$ \\
\bottomrule[0.65pt]
\end{tabular}  
\end{table*} 

\subsection{Photo-z Uncertainties}
\label{worse_z}

Photo-$z$ uncertainty is considered as a significant source of errors in weak lensing measurement. 
As our Quasi-2D method does use the galaxy redshifts, our results are likely to be sensitive to the errors in photo-$z$. In this section, we explore the effect of photo-$z$ on our parameter constraints. 

\begin{figure}[H]
\centering
\includegraphics[width=0.48\textwidth]{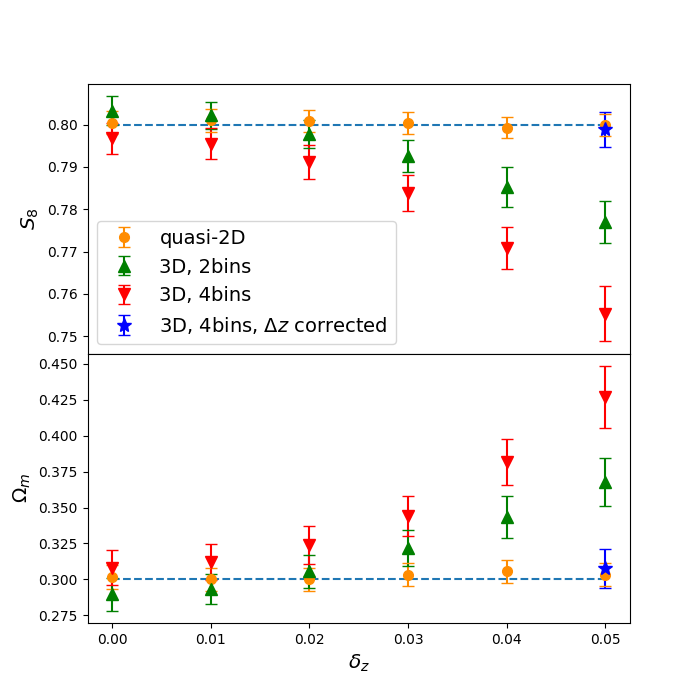}
\caption{The constraints of $S_8$ and $\Omega_{\rm m}$ under different assumptions about the photo-$z$ uncertainties in the mock data. The orange data points represent the results obtained from the quasi-2D analysis, while the green and red triangles correspond to the analysis with 2 and 4 redshift bins respectively.The blue pentagram represents the results corrected by the redshift bias $\Delta z$ in
each redshift bin.} \label{diffpz}
\end{figure}

Since we cannot know the true redshift distribution of the samples and their photo-z errors, in this section, we use mock data for investigation. We generate a set of random pairs following a correlated Gaussian distribution as simulated shear estimators on different redshifts, assuming a flat $\Lambda$CDM cosmology with $S_8=0.8$ and $\Omega_{\rm m}=0.3$. To increase the computational speed, we set the amplitude of the shape noise to be 0.08. We apply the same redshift distribution to the simulated galaxy pairs as for the actual data, and add a Gaussian random redshift error with a standard deviation of $\delta_z (1+z)$ to each galaxy. $\delta_z$ ranges from 0 to 0.05 to represent varying levels of photo-$z$ uncertainty, and Li et al. (in prep.) found that the errors for galaxies with  mag $\lesssim$ 21 in DECaLS are approximately $0.02(1+z)$. We analyse the simulated data using the same quasi-2D approach and ensure the same redshift cut. The resulting constraints with different photo-$z$ errors are shown by the orange dots in Figure \ref{diffpz}. Our analysis reveals that when using quasi-2D lensing, the cosmological constraints remain unbiased even at a substantial photo-$z$ error of $\delta_z=0.05$, demonstrating a certain degree of tolerance to photo-$z$ uncertainties. 

In Li et al.'s study (in prep.) on the photometric redshift uncertainty of our shear catalog, they note a potential redshift bias, $\Delta z_p= -0.08(z_s-0.57)$, where $z_s$ is the spectral redshift of the galaxy. This is due to a peak in galaxy distribution at the redshift about 0.5, as seen in Figure \ref{fig:zpdf}. There is a trend that samples with photo-$z$ lower than 0.5 come from a higher redshift, and those with higher photo-$z$ possibly originating from lower redshift. 
To assess the impact of this bias on the quasi-2D analysis, we incorporate such a redshift bias $\Delta z_p$ into our simulations. We find that the additional $\Delta z_p$ leads to negligible changes of the best-fit values of $S_8$ and $\Omega_m$, less than one thousandth in relative amplitudes.

\begin{figure}[H]
\centering
\includegraphics[width=0.45\textwidth]{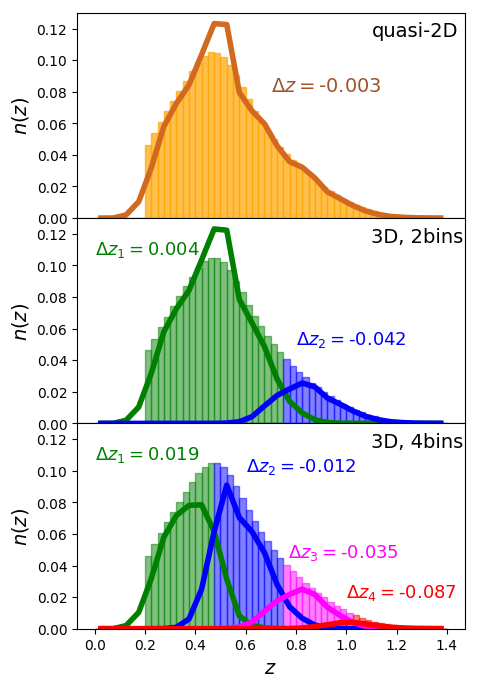}
\caption{The distributions of photo-$z$ in simulated data (histograms, with $\delta z = 0.05$) and the corresponding true redshift distributions (solid lines) within each redshift bin (different colors) under various analyses. $\Delta z_i$ represents the average true $z$ (solid lines) minus the average photo-$z$ (histograms) for galaxies in the $i$-th redshift bin, and their colors correspond to the colors of the redshift bins in the figure.} \label{fig:simlzpdf}
\end{figure}

We further divided the simulated data into 2 and 4 redshift bins. Within these bins, we conducted a quasi-2D analysis to measure auto- and cross-correlations and constrain the parameters, which can be denoted as a form of 3D analysis. The results are presented as upright and inverted triangle (green and red) data points in Figure \ref{diffpz}. Notably, we observe that larger photo-$z$ errors correspond to smaller $S_8$ values and larger $\Omega_{\rm m}$ values. Moreover as the number of redshift bins increases, this bias becomes more pronounced. To investigate the underlying causes of significant bias, we present in Figure \ref{fig:simlzpdf} the photo-$z$ distributions (histograms) under different analysis and the corresponding true redshift distributions (solid lines) within each redshift bin when $\delta_z = 0.05$. Here, $\Delta z_i$ denotes the average true $z$ minus the average photo-$z$ for galaxies in the $i$-th redshift bin. In quasi-2D analysis, the overall redshift bias caused by photo-$z$ errors is minimal due to the absence of redshift binning. However, in 3D analysis, the redshift bias on different sides of the redshift peak exhibits opposite signs, and its magnitude increases with the number of redshift bins. This leads to erroneous redshift predictions, causing biases in parameter results. Consequently, we apply the $\Delta z_i$ from the bottom panel of Figure \ref{fig:simlzpdf} to correct the photo-$z$ in the 3D analysis with 4 redshift bins, represented by the blue pentagrams in Figure \ref{diffpz}. The corrected results revert to an unbiased state, albeit with slightly larger errors compared to quasi-2D results. In summary, photo-$z$ uncertainties induce biases in the average redshift of the sample, with more pronounced effects in 3D analysis, while quasi-2D analysis is largely unaffected. As such, we favour the quasi-2D lensing approach, as it is theoretically unbiased and less vulnerable to the photo-$z$ uncertainties.

Finally, with the real shear catalog, we can test the robustness of our results by adding an additional Gaussian random value with $\sigma_z=0.02(1+z)$ on each photo-$z$ of galaxy. 
The numerical results of parameters using both $\xi_{\gamma_t \gamma_t}$ and $\xi_{\gamma_\times \gamma_\times}$ components are presented in the row labeled "worse photo-$z$" in Table \ref{tab:allresults}. We observe that these additional redshift errors lead to a slight increase in the uncertainties of $A_{\rm IA}$ and $S_8$ and the value of $A_{\rm IA}$ due to the inaccurate redshift information. However, overall, this has little impact on the results of the three parameters. This suggests that our method is robust against redshift uncertainties.

\section{2D lensing analysis}\label{sec5}

We have so far introduced a way of analysing the shear-shear 2PCF by symmetrizing the joint PDF of the shear estimators, as shown in eq.(\ref{sym1}). The involving galaxy pairs can be divided into different bins of angular separations, but not into different redshift bins like what is done in tomographic studies. On the other hand, the photo-z information is used in computing the prediction of $\hat{\xi}$ for each galaxy pair, as shown in eq.(\ref{sym2}), called quasi-2D method. The quasi-2D method seems to be less sensitive to the redshift uncertainties as shown in the last section, but its disadvantage is also quite obvious: the data points of the shear 2PCFs are not visible at all. As the values of the 2PCFs on different scales are important diagnostic of potential systematics in the measurement, it is important to find out whether there is a way to visualize the quality of the data points themselves. Based on these thoughts, we propose another pure 2D method. 

\subsection{Theory}
According to eq.(\ref{sym2}), the symmetry of the joint PDF is affected by the shear-shear correlation weighted by the distribution of the "B" factors and the slope of the PDF at different redshifts. If we use one correlation function $\hat{\xi}$ to symmetrize the joint PDF of, say $(G_1,G_1')$, in an angular bin, its theoretical interpretation should take into account the weights introduced by the variation of the PDF as a function of the redshift. Although this is doable in practice, we take a simpler approach here: 
we just use $2 \times 2$ bins for binning the joint PDF of shear estimators, in other words, the symmetry of the PDF is assessed by only counting the number of galaxy pairs in the four quadrants, or whether each shear estimator is greater than zero.

\begin{figure*}[ht!]
\centering
\includegraphics[width=0.94\textwidth]{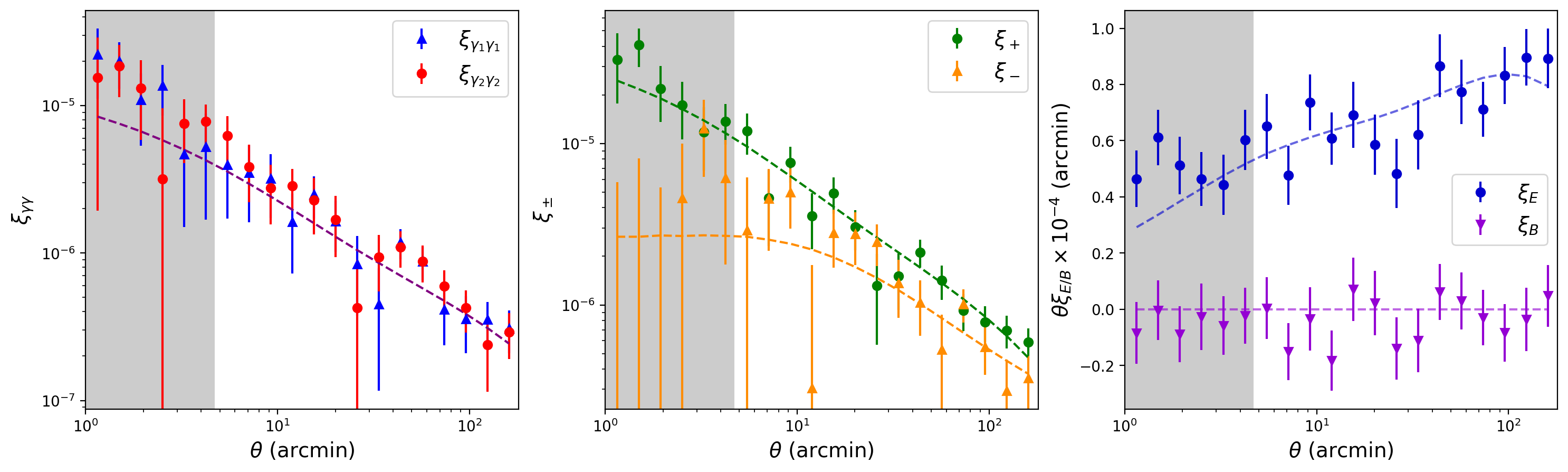}
\caption{The measured two-point correlation function $\xi_{\gamma_1 \gamma_1}$, $\xi_{\gamma_2 \gamma_2}$ (left panel) and $\xi_\pm$ (middle panel) using the pure 2D version of the PDF-SYM method, and the E/B-mode shear correlation functions in right panel. In the left panels, the purple dashed lines show the best-fit theoretical predictions using data of both $\xi_{\gamma_1 \gamma_1}$ and $\xi_{\gamma_2 \gamma_2}$ in the white region, with the numerical results shown in Table \ref{tab:allresults} labeled by "2D $\xi_{\gamma_1 \gamma_1} \& \xi_{\gamma_2 \gamma_2}$". In the middle and right panels, the dashed lines show the best-fit predictions of $\xi_\pm$, with the parameters shown in Table \ref{tab:allresults} labeled by "2D $\xi_\pm$". The data points in the grey areas are not used in the cosmological analysis.} \label{xipmdata}
\end{figure*}

\begin{figure}[H]
\centering
\includegraphics[width=0.43\textwidth]{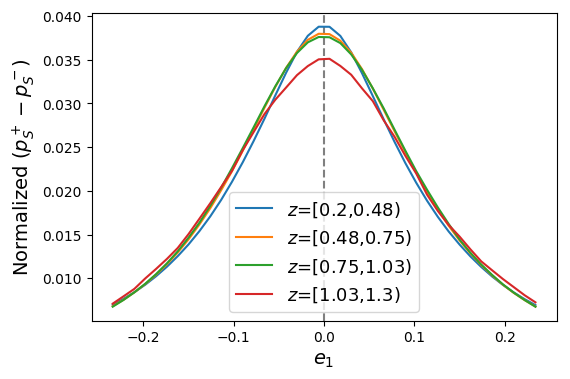}
\caption{The normalized probability distribution of $e_1$ ($\equiv G_1/|B|$) in 4 redshift bins. $p_S^+(e,z) -p_S^-(e,z)$ represents the distribution of galaxies with shear estimates $B$ greater than 0 subtracted from those of galaxies with $B < 0$. 
}
\label{pdfe1}
\end{figure}

Let us define a new observable $e_i \equiv G_i/|B|$, and modify eq.(\ref{Ghat}) slightly as: 
\begin{equation}\label{ehat}
\hat{e}_i = e_i-{\rm sgn}(B) \hat{g}_i = e_{S i}+{\rm sgn}(B)(g_i-\hat{g}_i),
\end{equation}
where $e_i$ is lensed quantity and $e_{S i}$ is the unlensed one. For convenience, in the rest of the discussion in this section, we neglect the subscript $i$. 
The number of galaxy pairs with both $\hat{e} > 0$ and $\hat{e}^{\prime} > 0$ is
\begin{eqnarray}
\label{nll1}
&&N_{>>}\\ \nonumber
&=&\int dz\int dz'\int_0^\infty d\hat{e}\int_0^\infty d\hat{e}'P(\hat{e},\hat{e}',z,z') \\ \nonumber
&=&\int dz\int dz'\int_0^\infty d\hat{e}\int_0^\infty d\hat{e}'P_S(e_S,e_S',z,z') 
\end{eqnarray}
where $P(\hat{e},\hat{e}',z,z')$ is the joint PDF of the new modified estimators, and $P_S(e_S,e_S',z,z')$ is the PDF of the unlensed quantities. Using eq.(\ref{ehat}), we can further get:
\begin{eqnarray}
\label{nll2}
&&N_{>>}\\ \nonumber
&=&\int dz\int dz' \int dg\int dg' \phi[g(z),g'(z')] \\ \nonumber
&\times&\int d\hat{g}\int d\hat{g}' \hat{\phi}(\hat{g},\hat{g}') \int_0^\infty d\hat{e} \int_0^\infty d\hat{e}'\cdot\sum_{s,s'}f(s)f(s') \\ \nonumber
&\times& P_S \big[\hat{e}+s(\hat{g}-g(z)),\hat{e}'+s'(\hat{g}'-g'(z')),z,z' \big],
\end{eqnarray}
in which $s$ and $s'$ stand for ${\rm sgn}(B)$ and ${\rm sgn}(B')$, which can only take the value of $1$ or $-1$. $f(s)$ and $f(s')$ are their corresponding frequencies of occurrence. $\phi(g,g')$ represents the PDF of real shear pairs, and $\hat{\phi}(\hat{g},\hat{g}')$ represents the PDF of pseudo-shear pairs. Using Taylor expansion, we have:
\begin{eqnarray}
&&P_S \big[\hat{e}+s(\hat{g}-g),\hat{e}'+s'(\hat{g}'-g'),z,z' \big]\\ \nonumber
&\approx& P_S(\hat{e},\hat{e}',z,z')+s(\hat{g}-g) \partial_{\hat{e}} P_S+s'(\hat{g}'-g') \partial_{\hat{e}'} P_S \\ \nonumber
&&+\frac{1}{2}(\hat{g}-g)^2 \partial_{\hat{e}} \partial_{\hat{e}} P_S + \frac{1}{2}(\hat{g}'-g')^2 \partial_{\hat{e}'} \partial_{\hat{e}'} P_S\\ \nonumber
&&+ss'(\hat{g}-g)(\hat{g}'-g') \partial_{\hat{e}} \partial_{\hat{e}'} P_S
\end{eqnarray}
We can similarly calculate the occupation number of the other PDF bins: $N_{<<}(\hat{e} < 0, \hat{e}' < 0)$, $N_{><}(\hat{e} > 0, \hat{e}' < 0)$, and $N_{<>}(\hat{e} < 0, \hat{e}' > 0)$. Due to the even parity of $P_S(e_S,e_S',z,z')$ with respect to both $e_S$ and $e_S'$, we can derive a very concise expression on the PDF symmetry as follows:
\begin{eqnarray}
&&N_{>>}+N_{<<}-N_{><}-N_{<>} \\ \nonumber
&=&4\int dz\int dz'\int dg\int dg' \phi[g(z),g'(z')] \\ \nonumber
&\times& \int d\hat{g}\int d\hat{g}' \hat{\phi}(\hat{g},\hat{g}') \int_0^\infty d\hat{e} \int_0^\infty d\hat{e}'\cdot\sum_{s,s'}f(s)f(s') \\ \nonumber
&\times& ss'(\hat{g}-g)(\hat{g}'-g') \partial_{\hat{e}} \partial_{\hat{e}'} P_S  \\ \nonumber
&=&4\int dz\int dz' [P_S^+(0,0,z,z') \\ \nonumber
&&-P_S^-(0,0,z,z')]  [\langle \hat{g}\hat{g}' \rangle +\langle g(z)g'(z')\rangle] 
\end{eqnarray}
where $P_S^+(0,0,z,z')$ denotes the probability at the zero point on the redshift $(z,z')$ with $ss' > 0$, and $P_S^-(0,0,z,z')$ denotes the same but with $ss' < 0$. For the joint PDF to reach a symmetric state in the case of $2\times 2$ bins, i.e., for $N_{>>}+N_{<<}=N_{><}+N_{<>}$, $\langle \hat{g}\hat{g}^\prime \rangle$ would be the estimator of total shear-shear correlation, which is shown as follows:
\begin{equation}
\label{ggfinal1}
\langle \hat{g}\hat{g}' \rangle=-\frac{\int dz\int dz' w(z,z')\langle g(z) g' (z')\rangle}{\int dz\int dz' w(z,z')}, 
\end{equation}
in which $w(z,z')$ is the weight given by the PDF shape:
\begin{equation}
\label{ggfinal2}
w(z,z')= P_S^+(0,0,z,z') -P_S^-(0,0,z,z').
\end{equation}
In practice, we find that as a good approximation, we can factorize $w(z,z')$ as:
\begin{equation}
w(z,z')\approx [p_S^+(0,z) -p_S^-(0,z)]  [p_S^+(0,z') -p_S^-(0,z')]
\end{equation}
Figure \ref{pdfe1} displays the normalized distribution of $p_S^+(e,z) -p_S^-(e,z)$ of the $e_1$ component in different redshift bins. These distribution functions are similar to Gaussian functions. The weights $w(z,z')$ in our 2D analysis are proportional to the number of galaxy pairs at different redshifts $n(z,z')$, and the inverse of the standard deviations of $e$ (which can be thought of as shape noise) at the two redshifts. It is worth noting that as redshift increases, the height of the normalized PDF decreases, indicating that the shape noise is more significant for galaxy pairs at larger redshifts. Our method therefore assigns lower weights to them. The situation of $e_2$ is similar, therefore not shown here.

\subsection{Results}

\begin{figure}[H]
\centering
\includegraphics[width=0.46\textwidth]{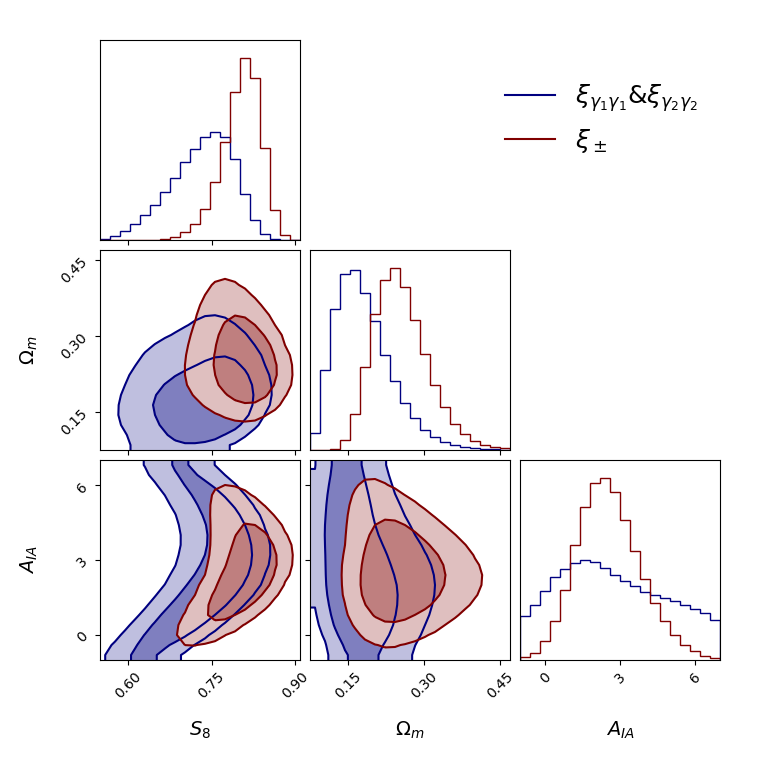}
\caption{The 68\% and 95\% confidence level contour plots of three parameters: $S_8$, $\Omega_{\rm m}$, and $A_{\rm IA}$, using the pure 2D analysis. Blue contours show the joint constraints from auto-correlation of $\gamma_1$ and $\gamma_2$, and red contours shows those from $\xi_\pm$. The data points of correlation functions are shown in Figure \ref{xipmdata}. Numerical results are listed in Table \ref{tab:allresults}. \label{xipmcontour}}
\end{figure}

Figure \ref{xipmdata} illustrates the two-point correlation functions $\xi_{\gamma_1 \gamma_1}$, $\xi_{\gamma_2 \gamma_2}$ and $\xi_\pm (\theta)$ from the pure 2D method. Note that the data on small scales with the grey color is excluded from the analysis, as the case in the quasi-2D analyses. We again employ the Jackknife approach to estimate the covariance matrix, and use the MCMC technique \citep{2000astro.ph..6401C} to constrain $S_8$, $\Omega_{\rm m}$ and $A_{\rm IA}$, as shown in Figure \ref{xipmcontour}. The specific numerical values are shown in Table \ref{tab:allresults}. The results obtained using $\xi_\pm$ are superior to those obtained using $\xi_{\gamma_1 \gamma_1}$ and $\xi_{\gamma_2 \gamma_2}$, although, they are consistent. In comparison to the quasi-2D analysis, the constraints by the 2D analysis are generally looser, which is mainly due to the insufficient use of the redshift information.

In \S\ref{internal}, we show that the quality of results using $\xi_{\gamma_1 \gamma_1}$ is inferior to that using $\xi_{\gamma_2 \gamma_2}$ due to systematic errors in measuring $\gamma_1$. In the 2D analysis, the results from $\xi_{\gamma_1 \gamma_1}$ in Table \ref{tab:allresults} suggest a notable underestimation of $\Omega_{\rm m}$, as also observed in the quasi-2D analyses. This discrepancy is also reflected in the left panel of Figure \ref{xipmdata}, where the blue points ($\xi_{\gamma_1 \gamma_1}$) exhibit a lower trend on small scales compared to the red points ($\xi_{\gamma_2 \gamma_2}$). Unlike the quasi-2D analysis, the error bars of parameters using $\xi_{\gamma_1 \gamma_1}$ and $\xi_{\gamma_2 \gamma_2}$ separately in the 2D analysis are similar. The key distinction between these two analyses is whether redshifts of galaxy pairs are utilized. Hence, we speculate that this disparity is attributed to redshift-dependent systematics, which are mitigated when combining results across all redshifts. 

Finally, we perform the E/B mode decomposition of our 2D correlation data as a way of checking the potential systematics. 
We employ the method proposed by Crittenden \cite{2002ApJ...568...20C}, in which the E/B mode shear correlation functions are given by
\begin{equation}
\xi_{E /B}(\theta)=\frac{\xi_{+}(\theta) \pm \xi^{\prime}(\theta)}{2}
\end{equation}
where "+" for E-mode and "-" for B-mode, and
\begin{equation}\label{xiprime}
\xi^{\prime}(\theta)=\xi_{-}(\theta)+4 \int_\theta^{\infty} \frac{d \phi}{\phi} \xi_{-}(\phi)-12 \theta^2 \int_\theta^{\infty} \frac{d \phi}{\phi^3}\xi_{-}(\phi)
\end{equation}
For the integral in eq.(\ref{xiprime}) from $\theta$ to infinity, we employ the best-fit predictions of $\xi_\pm$ in the 2D analysis for the range $\theta > \theta_{\rm max}$=180', where the integral is insensitive to the parameters \cite{2010A&A...516A..63S}. The measured $\xi_{E /B}(\theta)$ are shown in the right panel of Figure \ref{xipmdata}, and their theoretical predictions are present by the dashed lines. We observe no significant B-mode signal across all angular scales, and the E-mode aligns well with theoretical predictions. Furthermore, we compute the probability of the B-mode being zero within the range $4.7' < \theta < 180'$ is $p=0.54$ with $\chi^2=0.92$ per degree of freedom on average. Hence, we believe that our correlation function measurements show no noticeable impact from any B-mode contamination.

Given that our shear-shear correlation is measured only with galaxy images from two different exposures, the systematic errors from correlated PSF residuals should have been strongly suppressed \citep{Lu_2017}. The remaining systematics should be mainly due to instrumental issues, which is usually hard to remove in practice. However, given that our shear catalog keeps the positional information of each galaxy image on the CCD, it is possible to directly calibrate such a bias. Further examination of systematics will be presented in a separate work (Shen et al. in prep.).

\section{Dicussion and Conclusion}\label{sec6}
\begin{figure*}[ht!]
\centering
\includegraphics[width=0.69\textwidth]{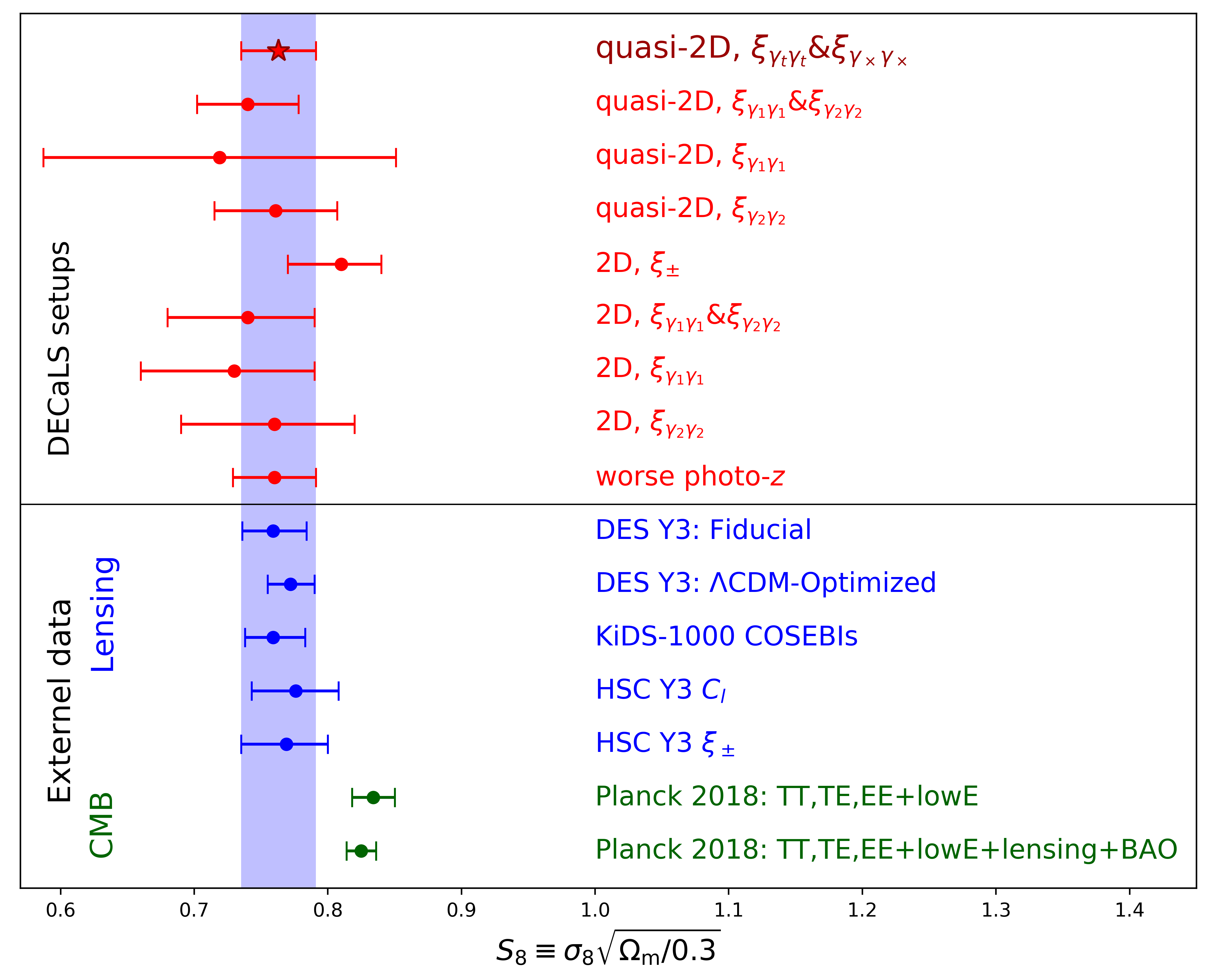}
\caption{Summary of the $S_8$ constraints from various consistency tests, as well as from other lensing surveys and Planck. \label{diff_S8}}
\end{figure*}

Figure \ref{diff_S8} summarizes our results of $S_8$ in different settings, and presents some $S_8$ constraints from other lensing surveys and Planck. The results of lensing surveys are from the fiducial and $\Lambda$CDM-optimized analyses in DES Y3~\citep{2022PhRvD.105b3514A}, KiDS-1000 cosmology~\citep{2021A&A...645A.104A}, the analyses of $C_l$ and $\xi_\pm$ in HSC Y3~\citep{2023arXiv230400702L, 2023arXiv230400701D}. Despite the fact that the DECaLS data are used to prepare for the spectroscopic survey of DESI and the quality of galaxy images is in general much worse than those of the typical weak lensing surveys, our constraints on $S_8$ are quite competitive and consistent with those from the other surveys. In the lower part of Figure \ref{diff_S8}, we present the $S_8$ constraints obtained from the Planck 2018 dataset~\citep{2020A&A...641A...6P} using the baseline TT, TE, EE+lowE likelihood, as well as datasets that incorporate CMB lensing and BAO. We observe that the Planck predictions for $S_8$ exceed our baseline constraints by about 2$\sigma$, as is the case for most of the other lensing surveys.

Overall, this paper proposes a new statistical approach for calculating the shear-shear 2PCF, i.e., the quasi-2D shear 2PCFs with the PDF-SYM method. The PDF-SYM method is very different from the common method of averaging the galaxy ellipticities. It is more close to finding the median of the shear estimators. We use the shear catalogue processed by the Fourier\_Quad method~\citep{2022arXiv220602434Z}. We select the galaxies in the redshift range of $[0.2, 1.3]$ and angular range from $\theta=4.7$ arcmin to 180 arcmin. Importantly, we use the photo-$z$ of each galaxy instead of dividing redshift bins in the quasi-2D analysis. We conduct multiple consistency tests and examine the impact of photo-$z$ to ensure the robustness of our results. We also derive the formalism for a pure 2D analysis of the shear-shear correlation, and place the corresponding cosmological constraints as well. Our main conclusions are as follows:

\begin{itemize}

\item We demonstrate the ability of our novel PDF-SYM method in analysing the quasi-2D shear 2PCFs. We choose the angular range conservatively to avoid the effects of some unknown physics at very small scales and obtain the constraints of $S_8=0.762 \pm 0.026$ and $\Omega_m=0.234 \pm 0.075$. In addition, despite not dividing redshift bins, our method successfully detects the effects of IA and determines a value of $A_{\rm IA}=1.59 \pm 0.73$.

\item We examine the robustness of different components of shear 2PCFs. Our analysis reveals that the results using $\xi_{\gamma1 \gamma1}$ are not as accurate as those using $\xi_{\gamma2 \gamma2}$, likely due to the additional shear uncertainties induced by the CCD electronics in the direction of $\gamma_1$.

\item We use mock data to test the robustness of our analysis against the photo-z uncertainties. Our findings indicate that even with redshift errors up to $0.05(1+z)$, the constraints on the parameters remain highly stable. On the contrary, if redshift is binned, the photo-z errors will introduce bias to the average redshift within each redshift bin, leading to biases in the constraints on parameters.

\item We find that the 2D shear correlation functions in the PDF-SYM method are effectively weighted differently from conventional approaches. Specifically, the weights are not only proportional to the number of galaxies at different redshifts but also inversely proportional to the shape noise. However, compared to quasi-2D analysis, 2D analysis insufficiently use the redshift information of each galaxy, leading to somewhat less accurate parameter constraints. 

\end{itemize}

Our work demonstrates the power of the PDF-SYM method in analysing the quasi-2D shear 2PCFs. It is a valuable consistency test of the weak lensing measurement of the cosmological parameters, because the PDF-SYM algorithm is quite different from the usual weighted-averaging method in several aspects: 1. it is less affected by the potential outliers; 2. it allows direct use of the unnormalized forms of, e.g., the quadrupole moments (defined in either Fourier or real space), providing more options in dealing with various systematic effects, such as the photon noise as shown in \cite{2017ApJ...834....8Z}; 3. it automatically minimizes the statistical uncertainty of the result without requiring weightings of the shear estimators. In principle, the PDF-SYM algorithm can also be applied to other shear estimators such as the usual galaxy ellipticities. To do so, one should not only take into account the multiplicative and additive biases (which is quite straightforward), but also the parity properties of the shear responsivities, as shown in Ref.~\cite{2017ApJ...834....8Z} for the FQ shear estimators.

The quasi-2D analysis can be applied to scenarios where signals from different redshifts are combined to estimate cosmological parameters with relatively low contamination from photo-z errors. In the near future, many upcoming large scale surveys will provide a large number of images of faint galaxies, which will greatly leverage the capabilities of the FQ method and the PDF-SYM method. In a companion paper (Shen et al., in prep.), we will carry out a tomographic study of the shear-shear correlation, in which more specific discussions about redshift-dependent systematic uncertainties will be presented.

\textbf{Acknowledgements} This work is supported by the National Key Basic Research and Development Program of China (2023YFA1607800, 2023YFA1607802), the NSFC grants
(11621303, 11890691, 12073017), and the science research grants from China Manned Space Project (No.CMS-CSST-2021-A01). The computations in this paper were run on the $\pi$2.0 cluster supported by the Center for High Performance Computing at Shanghai Jiao Tong University. The Legacy Imaging Surveys of the DESI footprint is supported by the Director, Office of Science, Office of High Energy Physics of the U.S. Department of Energy under Contract No. DE-AC02-05CH1123, by the National Energy Research Scientific Computing Center, a DOE Office of Science User Facility under the same contract; and by the U.S. National Science Foundation, Division of Astronomical Sciences under Contract No. AST-0950945 to NOAO. The Photometric Redshifts for the Legacy Surveys (PRLS) catalog used in this paper was produced thanks to funding from the U.S. Department of Energy Office of Science, Office of High Energy Physics via grant DE-SC0007914.

\bibliography{sample631}
\bibliographystyle{abbrv}

\end{multicols}
\end{document}